\documentclass[pre,a4paper,showpacs,byrevtex]{revtex4}
\usepackage{graphicx}
\usepackage{dcolumn}
\usepackage{amsmath}
\usepackage{array}
\usepackage{bm}
\usepackage{amssymb}
\usepackage{amsfonts}

\begin{document}

\title{Lagrange-mesh calculations and Fourier transform}

\author{Gwendolyn Lacroix}
\email[E-mail: ]{gwendolyn.lacroix@umons.ac.be}
\author{Claude Semay}
\email[E-mail: ]{claude.semay@umons.ac.be}
\affiliation{Service de Physique Nucl\'{e}aire et Subnucl\'{e}aire,
Universit\'{e} de Mons - UMONS, 
Acad\'{e}mie universitaire Wallonie-Bruxelles, 
Place du Parc 20, 7000 Mons, Belgium }

\date{\today}

\begin{abstract}
The Lagrange-mesh method is a very accurate procedure to compute 
eigenvalues and eigenfunctions of a two-body quantum equation.
The method requires only the
evaluation of the potential at some mesh points in the configuration space.
It is shown that the eigenfunctions can be easily computed in the
momentum space by a Fourier transform using the properties of the basis functions. 
Observables in this
space can also be easily obtained.
\end{abstract}

\maketitle

\section{Introduction}
\label{sec:intro} 

The Lagrange-mesh method is a very accurate procedure to
compute eigenvalues and eigenfunctions of a two-body Schr\"{o}dinger
equation \cite{BH-86,VMB93,Ba-95,Baye06,baye08} as well as a  
semirelativistic Hamiltonian \cite{sema01,brau02,buis05,lag1}.
The trial eigenstates are developed in a basis of well chosen functions,
the Lagrange functions. Using their special properties, the potential 
matrix elements are simply the values of the potential at
mesh points, if they are computed with a Gauss quadrature.
At first sight, this method could look like a discrete variational method,
but this is absolutely not the case since the eigenfunctions can be
computed at any position. Because of the use of the Gauss quadrature
scheme, the method is not variational but a great accuracy can nevertheless
be reached \cite{baye02}. The method presented here relies on a mesh of points 
built with the zeros of a Laguerre polynomial, but a general procedure
for deriving other Lagrange meshes related to orthogonal or non-orthogonal 
bases has also been developed \cite{baye99}. Even if we only focus on two-body
systems in this paper, it is worth mentioning that this method can be extended 
to treat very accurately three-body systems as well in nuclear physics 
as in atomic physics (see for instance Ref.~\cite{hess99}).

At the beginning, this method was developed in the position space. As we will
see below, the potential matrix elements are very easy to compute if the interaction 
is known in terms of the distance $r$ between the interacting particles. This is also 
true for mean values of observables depending on $r$. For some problems, 
it can be also useful to compute the eigenfunctions in the momentum space
by the Fourier transform, as well as 
observables depending on the relative momentum between the particles. We will show that
the Lagrange-mesh method can provide these type of data very efficiently
and very easily, using the fundamental properties of the Lagrange functions.

The Lagrange-mesh methods in configuration space is
described in Sec.~\ref{sec:method}, while Sec.~\ref{sec:momspace} presents 
some results in momentum space. 
An ansatz to compute easily the only non-linear parameter of the method
is described in Sec.~\ref{sec:scale}.
Test calculations are presented in Sec.~\ref{sec:num}, and some
concluding remarks are given in Sec.~\ref{sec:crem}.

\section{Method in position space}
\label{sec:method}

\subsection{Lagrange functions}
\label{ssec:Lagfunc}

The basic ingredients for the Lagrange-mesh method are a mesh of 
$N$ points $x_i$ associated with an orthonormal set of $N$ indefinitely 
derivable functions $f_j (x)$ \cite{BH-86,VMB93,Ba-95}.
The Lagrange function $f_j(x)$ satisfies the Lagrange conditions, 
\begin{equation}
\label{2.1}
f_j (x_i) = \lambda_i^{-1/2} \delta_{ij},
\end{equation}
that is to say it
vanishes at all mesh points except one.
 The $x_i$ and $\lambda_i$ are respectively the abscissae and the weights of a Gauss
quadrature formula
\begin{equation}
\label{2.2}
\int^{\infty}_0 g(x) dx \approx \sum_{k=1}^{N} \lambda_k g(x_k).
\end{equation}
As we work with the radial part of wavefunctions, we consider 
the case of the Gauss-Laguerre quadrature because the domain 
of interest is $[0,\infty]$.
The Gauss formula (\ref{2.2}) is exact when $g(x)$ is a polynomial
of degree $2N-1$ at most, multiplied by $\exp (-x)$.
The Lagrange-Laguerre mesh is then based on the zeros of a Laguerre
polynomial of degree $N$ \cite{BH-86} and the mesh points are given by 
$L_N (x_i)=0$. These zeros can be determined with a high precision with usual methods 
to find the roots of a polynomial \cite{numrec} (the \emph{Mathematica} expression \texttt{Root} 
does the job efficiently) or as the eigenvalues 
of a particular tridiagonal matrix \cite{golu69}.
The weights can be computed by the following formula \cite{baye02}
\begin{equation}
\label{2.3}
\ln \lambda_i = x_i - \ln x_i + 2 \ln\Gamma(N+1)-\sum_{j\ne i=1}^N \ln(x_i-x_j)^2.
\end{equation}
It is worth noting that, for most calculations, it is not necessary 
to compute the weights $\lambda_i$. 
The original Lagrange functions do not vanish at origin, so 
it is preferable to use the regularized Lagrange functions whose explicit form 
is given by
\begin{equation}
\label{2.4}
f_i (x) = (-1)^i x_i^{-1/2} x (x-x_i)^{-1} L_N (x) \exp (-x/2),
\end{equation}
which is a polynomial of degree $N$, multiplied by an exponential function.
Such a function $f_i (x)$ vanishes at the origin and at $x_j$ with $j \neq i$.

With the Lagrange-mesh method, the solution of a quantum equation reduces 
(as it is often the case) to 
the determination of eigensolutions of a given matrix. Let us consider 
the eigenvalue equation
\begin{equation}
\label{2.5}
\left[ T(\vec p\,^2) + V(r) \right] |\psi\rangle = E\, |\psi \rangle,
\end{equation}
where $T(\vec p\,^2)$ is the kinetic energy term of the Hamiltonian and
$V(r)$ the potential which depends only on the radial coordinate $r=|\vec r\,|$.
In the following, we will always work in natural units: $\hbar = c = 1$.
A trial state $|\psi \rangle$, approximation of the genuine eigenstate, 
is expanded on a basis built with these regularized Lagrange functions
\begin{equation}
\label{2.6}
|\psi \rangle = \sum_{j=1}^N C_j |f_j \rangle \quad \textrm{with} \quad 
\langle \vec r\,|f_j \rangle = \frac{f_j (r/h)}{\sqrt{h}r} Y_{l m}
(\hat r),
\end{equation}
with $\hat r= \vec r/r$.
The coefficients $C_j$ are linear variational parameters and the scale
factor $h$ is a non-linear parameter aimed at adjusting the mesh to the
domain of physical interest.
Contrary to some other mesh methods, the wavefunction is also
defined between mesh points by (\ref{2.4}) and (\ref{2.6}).

Basis states $|f_i\rangle$ built with the regularized Lagrange functions 
are not exactly orthogonal. But, at the Gauss approximation, we have
$\langle f_j|f_i\rangle = \delta_{ji}$. So, in the following, all 
mean values will be performed using the Gauss quadrature formula (\ref{2.2}). 
In this case, the potential matrix elements are given by
\begin{equation}
\label{2.10}
\langle f_i|V(r)|f_j\rangle = V(h x_i)\,\delta_{ij}.
\end{equation}
The potential matrix is both simple to obtain and diagonal. Let us
assume that the matrix elements
$\langle f_i|T|f_j\rangle \approx T_{ij}$ are known.
Their computation will be explained in the next section. 
With (\ref{2.6}) and (\ref{2.10}), the variational
method applied to (\ref{2.5}) provides a system of $N$ mesh
equations
\begin{equation}
\label{2.11}
\sum_{j=1}^N\, [T_{ij} + V(h x_i)\, \delta_{ij} - E\, \delta_{ij}]\, C_j
= 0.
\end{equation}

In the Lagrange-mesh method, the Hamiltonian matrix elements are not exactly calculated,
but are computed at the Gauss approximation. So, the variational character of the
method cannot be guaranteed, except if an exact
quadrature is performed. In practice, for a sufficiently high number of
basis states, the method is often variational (eigenvalues computed are
all upper bounds) or antivariational (eigenvalues computed are all lower
bounds). It has been observed \cite{BH-86,VMB93,Ba-95} that the accuracy of
the mesh approximation remains close to the accuracy of the original
variational calculation without the Gauss approximation. So, in most cases,
a very high accuracy can be achieved in the framework of the Gauss approximation, 
though the mathematical reasons for the high efficiency of this method are not 
well known yet \cite{baye02}.

The accuracy of the eigensolutions depends on two
parameters: The number of mesh points $N$ and the value of the scale
parameter $h$. For a sufficiently high value of $N$ (which can be as low as 20 or 30),
the eigenvalues present a large plateau as a function of $h$. This is a
great advantage for the Lagrange-mesh method since the non-linear parameter must not
be determined with a high precision. Nevertheless, if $h$ is too small, a significant 
part of the wavefunction is not covered by the points of the Lagrange mesh.
When $h$ is too large, all points of the mesh are located in the 
asymptotic tail of the wavefunctions and it is then impossible to 
obtain good eigenvalues.  
So, it is interesting to have a procedure to estimate
directly a reasonable value of $h$ in order to avoid a search, which is always
time consuming. We have remarked that the best results are obtained when
the last mesh points are located ``not too far" in the asymptotic tail. So, if we choose a point
$r_\textrm{max}$ in the tail of the wavefunction, the value of $h$ can be obtained
by $h=r_\textrm{max}/x_N$, where $x_N$ is the last mesh point. A procedure
to estimate $r_\textrm{max}$ will be presented in Sec.~\ref{sec:scale}. 

\subsection{Kinetic parts}
\label{ssec:kp}

Let us first look at the matrix $P$ whose elements are $P_{ij}=\langle f_i|\vec p\,^2|f_j\rangle$.  
With (\ref{2.2}), these matrix elements are given by
\begin{equation}
\label{2.8}
P_{ij} = \frac{1}{h^2} \left( t_{ij} + \frac{l(l+1)}{x_i^2}
\delta_{ij}
\right),
\end{equation}
where $l$ is the orbital angular momentum quantum number, and where
\begin{equation}
\label{2.8bis}
t_{ij} = \int^{\infty}_0 f_i (x) \left( -\frac{d^2}{dx^2} \right)
f_j (x)\ dx \approx - \lambda_i^{1/2} f_j'' (x_i).
\end{equation}
This compact expression is exact for some Lagrange meshes.
This is not the case for the regularized Laguerre mesh.
An exact expression can easily be obtained
(see appendix in Ref.~\cite{VMB93}).
However, as shown in  Ref.~\cite{Ba-95}, it is
preferable
to use the approximation (\ref{2.8})-(\ref{2.8bis}).
The kinetic matrix elements are then even easier to obtain and read
\cite{Ba-95}
\begin{equation}
\label{2.9}
t_{ij} = \left \{ \begin{array}{lc}
(-)^{i-j} (x_i x_j)^{-1/2} (x_i+x_j)(x_i-x_j)^{-2}
& (i \neq j), \\
(12x_i^2)^{-1} [4 + (4N + 2) x_i - x_i^2]
& (i = j). \end{array} \right.
\end{equation}

For a nonrelativistic Hamiltonian, $T_{ij}= \frac{1}{2 \mu} P_{ij}$,
where $\mu$ is the reduced mass of the system. For a more general operator $T(\vec p\,^2)$, 
as the kinetic part of a spinless Salpeter equation
$2 \sqrt{\vec p\,^2 + m^2}$, the calculation is much more
involved. The idea is to use a four-step method suggested in Ref.~\cite{fulc94} 
(see also references therein) and applied in Ref.~\cite{sema01}:
\begin{enumerate}
\item Computation of the matrix $P$ whose elements are $P_{ij}=\langle f_i|\vec p\,^2|f_j\rangle$,
given by (\ref{2.8})-(\ref{2.9}).
\item Diagonalization of the matrix $P$. If $P^D$ is the diagonal
matrix formed by the eigenvalues of $P$, we have
\begin{equation}
\label{step2}
P = S\,P^D\,S^{-1},
\end{equation}
where $S$ is the transformation matrix composed of the normalized
eigenvectors.
\item Computation of $T^D$, a diagonal matrix obtained by
taking the function $T(x)$ of all diagonal elements of $P^D$
(For instance, $T(x)=2 \sqrt{x + m^2}$ for the case of a spinless Salpeter equation).
\item Determination of the kinetic matrix $T$ in the original basis
by using the transformation~(\ref{step2})
\begin{equation}
\label{step4}
T = S\,T^D\,S^{-1}.
\end{equation}
\end{enumerate}
The elements $T_{ij}$ of the matrix computed with (\ref{step4}) are
approximations of the numbers
$\langle f_i|T(\vec p\,^2)|f_j\rangle$. The calculation is not
exact for two reasons. First, the elements $T_{ij}$ are computed
with an approximate formula (\ref{2.8})-(\ref{2.9}). Second, the
diagonalization is performed in the limited definition space of the trial
function~(\ref{2.6}). In order to compute exactly the matrix elements of
the operator $T(\vec p\,^2)$, it is necessary to compute exactly
all eigenvalues of the infinite matrix whose elements are
$\langle T(\vec p\,^2) \rangle$, again exactly computed. This is
obviously not possible. It has been shown in Ref.~\cite{sema01}, that this
four-step procedure can give very good results.

\subsection{Mean values of radial observables}
\label{ssec:meanr}

The mean value of the operator $U(r)$ for a trial state $|\psi \rangle$ is given by
\begin{equation}
\label{meanU1}
\langle\psi |U(r)|\psi \rangle = \sum_{i,j=1}^N C_i\, C_j\,\langle f_i |U(r)|f_j \rangle.
\end{equation}
Using the Lagrange condition (\ref{2.1}) and the Gauss quadrature (\ref{2.2}), this integral reduces to
\begin{equation}
\label{meanU2}
\langle\psi |U(r)|\psi \rangle = \sum_{j=1}^N C_j^2\, U(h x_j).
\end{equation}
If $U$ is the identity, we recover the normalization condition as expected. A very high accuracy
can be obtained with this simple procedure \cite{baye08,hess99}.

\section{Method in momentum space}
\label{sec:momspace}

\subsection{Fourier transform}
\label{ssec:FT}

For some particular problems, it can be useful to compute the Fourier transform of a 
wavefunction in the position space in order to obtain the corresponding wavefunction 
in the momentum space. The Fourier transform $\phi^\textrm{FT}(\vec p\,)$ 
of a wavefunction $\phi(\vec r\,)$ is defined by
\begin{equation}
\label{FT1}
\phi^\textrm{FT}(\vec p\,)= \frac{1}{(2 \pi)^{3/2}}\int \phi(\vec r\,)\, e^{-i \vec p . \vec r} d\vec r.
\end{equation}
Using the spherical representation of the wavefunction
\begin{equation}
\label{FT2}
\phi(\vec r\,)= R_{n l}(r)\, Y_{lm}(\hat r),
\end{equation}
and using the spherical expansion of the function $e^{-i \vec p . \vec r}$ \cite{var}, it can be shown that
\begin{equation}
\label{FT3}
\phi^\textrm{FT}(\vec p\,)= R^\textrm{FT}_{n l}(p)\, \tilde Y_{l m}(\hat p),
\end{equation}
where $p=|\vec p\,|$ and $\hat p=\vec p/p$, and where
\begin{align}
\label{FT4}
R^\textrm{FT}_{n l}(p) &= (-1)^l \sqrt{\frac{2}{\pi}} \int_0^\infty R_{n l}(r)\, j_l(p\, r)\, r^2\, dr, \\
\label{FT5}
\tilde Y_{l m}(\hat p) &= i^l\, Y_{l m}(\hat p).
\end{align}
$j_l(x)$ is a spherical Bessel function \cite{abra65} 
and $\tilde Y_{l m}(\hat x)$ is called a modified spherical harmonic \cite{var}. 

Using expansion (\ref{2.6}), the radial part $R(r)$ of the trial function is given by
\begin{equation}
\label{FT6}
R(r) = \sum_{j=1}^N C_j \frac{f_j (r/h)}{\sqrt{h}r}.
\end{equation}
The Fourier transform $R^\textrm{FT}(p)$ of this radial function is defined by (\ref{FT4}). 
It is tempting to use the Gauss quadrature rule (\ref{2.2}) with the Lagrange condition (\ref{2.1})
to perform this calculation. The problem is that
spherical Bessel functions are rapidly oscillating functions. It is then not obvious that such a 
procedure could work. Actually, we have checked that the Fourier transform of a unique 
regularized Lagrange function, which is also a rapidly oscillating function, cannot 
be obtained in this way with a good accuracy. Fortunately, the radial part of a wavefunction has a much smoother behavior. As we will see on several examples in Sec.~\ref{sec:num}, its Fourier transform can be easily obtained in the framework of the Lagrange-mesh method by taking benefit of the very special properties
of the regularized Lagrange function. Using (\ref{2.2}) with (\ref{2.1}), 
the integral (\ref{FT4}) simply reduces to
\begin{equation}
\label{FT7}
\bar R^\textrm{FT}(p) = (-1)^l \sqrt{\frac{2}{\pi}} \, h^{3/2} \sum_{i=1}^N C_i\, \sqrt{\lambda_i}\, 
x_i\, j_l(h\, x_i\, p),
\end{equation}
where we use the ``bar" to indicate that this is not the exact Fourier transform $R^\textrm{FT}(p)$.
For a sufficiently high value of $N$ (which can be as low as 50), 
$\bar R^\textrm{FT}(p)\, \tilde Y_{l m}(\hat p)$
can be a very good approximation of the genuine eigenstate in the momentum space for values 
of $p \in [0,p_\textrm{max}]$, where $p_\textrm{max}$ can be determined with the 
procedure used to compute $r_\textrm{max}$ (see Sec.~\ref{sec:scale}). For values 
of $p \gtrsim p_\textrm{max}$, $\bar R^\textrm{FT}(p)$ can present large unphysical rapid oscillations. 
These oscillations do not develop in $R(r)$, because they are killed by the rapid 
decreasing of the regularized Lagrange functions. 

\subsection{Mean values of momentum dependent observables}
\label{ssec:meanp}

The mean value of the operator $K(p)$ for a trial states $|\psi \rangle$ is given by
\begin{equation}
\label{meanK1}
\langle\psi |K(p)|\psi \rangle = 
\int_0^\infty K(p)\, \left(R^\textrm{FT}(p)\right)^2\, p^2\, dp,
\end{equation}
where the angular part is already integrated. 
In this formula, the function $R^\textrm{FT}(p)$ can be replaced by $\bar R^\textrm{FT}(p)$.
Good results can sometimes be obtained, but the accuracy cannot be always guaranteed. This is
the case when the observable grows rapidly with $p$ and needs a very good quality of the
asymptotic tail of the wavefunction in the momentum space. Actually, it is easier and 
much more efficient to compute directly
\begin{equation}
\label{meanK2}
\langle\psi |K(p)|\psi \rangle = \sum_{i,j=1}^N C_i\, C_j\, \langle f_i |K(p)|f_j \rangle.
\end{equation}
The matrix elements $\langle f_i |K(p)|f_j \rangle$ can be determined by a procedure identical 
to the one used to compute $\langle f_i |T(\vec p\,^2)|f_j \rangle$. An intermediate step 
is the calculation of the matrix $K^D$, a diagonal matrix obtained by
taking the function $K(\sqrt{x})$ of all diagonal elements of $P^D$ 
(remember that $P$ is linked to the matrix elements of $\vec p\,^2$, not $p$). 
The numbers $\langle f_i |K(p)|f_j \rangle$
are well approximated by the elements of the matrix $K$ obtained by using the 
transformation~(\ref{step2}): $K = S\,K^D\,S^{-1}$. As we will see below, a very good accuracy can be
reached for the mean values $\langle K(p) \rangle$.

\section{Scale parameter}
\label{sec:scale}

An estimation of $r_\textrm{max}$ can be computed using the technique developed in
Ref.~\cite{brau98}. The first step is to find a potential $V_\infty (r)$
which matches at best the potential $V(r)$ for $r \rightarrow \infty$.
Three cases are considered in Ref.~\cite{brau98}: 
\begin{itemize}
\item $\kappa\, r^p$ with $\kappa >0$ and $p >0$; 
\item $-\kappa/r^p$ with $\kappa >0$ and $0 <p \le 1$;  
\item a square well.
\end{itemize}
The second step is to choose a trial state $| \lambda \rangle$ which
depends on one parameter $\lambda$, taken as the inverse of
a distance. Two cases are considered in Ref.~\cite{brau98}:
$u_\lambda(r)\propto r^{l+1}\, e^{-\lambda^2 r^2/2}$ (harmonic oscillator state) and
$u_\lambda(r)\propto r^{l+1}\, e^{-\lambda\,r}$ (hydrogen-like state), depending on $V_\infty (r)$.
If the quantum number $n$ is not zero, an effective value of $l$ is used
(see Ref.~\cite{brau98}). In a third step, 
the optimal value of $\lambda$ is determined by the usual condition
\begin{equation}
\label{lamb}
\frac{\partial}{\partial \lambda} \langle \lambda | T + V_\infty(r)  | \lambda
\rangle = 0,
\end{equation}
where $T$ is the kinetic part of the Hamiltonian considered. In the
case of complicated $T$ function, the following approximation can be used
\begin{equation}
\label{appT}
\left\langle T(\vec p\,^2) \right\rangle \rightarrow 
T\left(\langle \vec p\,^2\rangle\right).
\end{equation}
In particular, we have 
\begin{equation}
\label{ineq}
\left\langle \sqrt{\vec p\,^2 + m^2} \right\rangle \leq
\sqrt{\langle \vec p\,^2 \rangle + m^2 }.
\end{equation}
Various expressions for the optimal parameter $\lambda$ are given in Ref.~\cite{brau98}.

Introducing the dimensionless variable $s=\lambda\, r$, 
the regularized radial part $u_\lambda(s)$ of the trial state
$| \lambda \rangle$
is then analyzed to find the value of $s_\epsilon$ which satisfies the
following condition
\begin{equation}
\label{eps}
\frac{u_\lambda(s_\epsilon)}{\max_{s\in [0,\infty]} \left[ u_\lambda(s)
\right]} = \epsilon,
\end{equation}
where $\epsilon$ (typically in the range $10^{-4}$-$10^{-8}$) 
is a number small enough to neglect the contribution of
$u_\lambda(s)$ for values of $s$ greater than $s_\epsilon$. This is the last step of 
the procedure, which is very fast and whose details are given in Ref.~\cite{brau98}.
Note that equation (36) in Ref.~\cite{brau98} has an analytical solution given by
($x_N$ is replaced here by $s_\epsilon$ in order to match 
the present notations and to avoid a confusion with the last Lagrange-mesh point)
\begin{equation}
\label{xeps}
s_\epsilon = \left[-(l+1) W_{-1}\left( - \frac{\epsilon^{m/(l+1)}}{e} \right)  \right]^{1/m},
\end{equation}
where $W_{-1}$ is the Lambert function \cite{LambW} and $m=1$ or 2 depending on the trial function
$u_\lambda(r)$. 

At this stage, the ratio $s_\epsilon/\lambda$ corresponds approximately to a radial distance 
in the asymptotic tail of an eigenstate of the Hamiltonian $T + V_\infty(r)$. The idea is to 
identify this distance with the value of $r_\textrm{max}$ for the genuine Hamiltonian 
considered. It has been shown in Ref.~\cite{sema01} that this procedure works quite well and 
can give a value of the scale parameter $h$ ($h=r_\textrm{max}/x_N$) in the plateau 
mentioned above. The efficiency of this ansatz is due to the fact that the value of $h$
must not be known with a great accuracy in the Lagrange-mesh method. So, a crude determination of 
$r_\textrm{max}$ is sufficient and it is not necessary to go beyond the use of the very simple trial 
functions $u_\lambda(r)$ mentioned above and the approximation (\ref{appT}) for the computation
of the kinetic contribution.

To determine an estimation of $p_\textrm{max}$, let us look at the Fourier transform 
$u^\textrm{FT}_\lambda(s=p/\lambda)$ 
of the trial states considered $u_\lambda(s=\lambda\, r)$:
\begin{alignat}{3}
\label{u1}
&u_\lambda(s)\propto s^{l+1}\, e^{-s^2/2} &\quad \Rightarrow\quad  &u^\textrm{FT}_\lambda(s) \propto s^{l+1}\, e^{-s^2/2}, \\
\label{u2}
&u_\lambda(s)\propto s^{l+1}\, e^{-s} &\quad \Rightarrow\quad  &u^\textrm{FT}_\lambda(s) \propto \frac{s^{l+1}}{(s^2+1)^{l+2}}.
\end{alignat}
If $u_\lambda(s)$ is a harmonic oscillator state,
$u^\textrm{FT}_\lambda(s)$ has the same form. So it seems quite natural to set
$p_\textrm{max}=\lambda\, s_\epsilon$, since both functions present the same ratio (\ref{eps}) at the same value of their dimensionless argument. If the trial state is a hydrogen-like state, the situation is different 
since $u^\textrm{FT}_\lambda(s)$ decreases much more faster than $u_\lambda(s)$ for large (but not too large) values of $s$. Nevertheless, the simple choice $p_\textrm{max}=\lambda\, s_\epsilon$ works quite well also, as it will be shown below. So, finally, we have
\begin{equation}
\label{max}
r_\textrm{max}=s_\epsilon/\lambda \quad \textrm{and} \quad p_\textrm{max}=\lambda\, s_\epsilon,
\end{equation}
with $s_\epsilon$ and $\lambda$ determined by the procedure described above. 

\section{Numerical tests}
\label{sec:num}

In this section, several tests will be performed for the Lagrange-mesh method 
with both nonrelativistic and semirelativistic kinematics.
We will focus on the quality of wavefunctions and observables in the momentum space
since the efficiency of the method in the position space has already been 
demonstrated elsewhere \cite{BH-86,VMB93,Ba-95,Baye06,baye08,sema01,brau02,buis05,lag1}.
In order to estimate more precisely the quality of the Fourier transform (\ref{FT7}), we define a 
``quality factor" $Q(p_*)$ 
\begin{equation}
\label{QF}
Q(p_*)= \max_{p\in[0,p_*]}\left|  \frac{\bar u^\textrm{FT}(p) - u^\textrm{FT}(p)}{\max_{p\in[0,p_*]}| 
u^\textrm{FT}(p) |} \right|,
\end{equation}
where $\bar u^\textrm{FT}(p)/p = \bar R^\textrm{FT}_{n l}(p)$ given by (\ref{FT7}) and 
$u^\textrm{FT}(p)/p= R^\textrm{FT}(p)$ 
is the exact solution in momentum space. 

\subsection{Confining semirelativistic Hamiltonian}
\label{sec:num1}

Let us consider the ultrarelativistic two-body system with a quadratic potential 
\begin{equation}
\label{HUR}
H=2 \sqrt{\vec p\,^2} + a \, r^2.
\end{equation}
This Hamiltonian is particularly interesting because it is probably the only one 
with a semirelativistic kinematics which is (partly) analytically solvable. With
an appropriate change of variable, this Hamiltonian can be recast into the form of 
a nonrelativistic Hamiltonian with a linear interaction \cite{luch91}, for which
solutions are known for S-states. The eigenvalues for $l=0$ are given by
\begin{equation}
\label{EUR}
E_{n0}=(4 a)^{1/3} |\alpha_n|,
\end{equation}
where $\alpha_n$ is the $(n+1)$th zero of the Airy function Ai  \cite{abra65}.
The corresponding regularized eigenfunctions are obtained directly in the momentum space 
\cite{Sem03}
\begin{equation}
\label{foUR}
u^\textrm{FT}_{n0}(p)= p R_{n0}(p) = \frac{1}{\textrm{Ai}'(\alpha_n)} \left( \frac{2}{a} \right)^{1/6}
\textrm{Ai}\left(\left( \frac{2}{a} \right)^{1/3}p + \alpha_n\right).
\end{equation}
Let us note that $\int_{\alpha_n}^\infty \textrm{Ai}^2 (s)\, ds = {\textrm{Ai}'}^2(\alpha_n)$.
Using the generalized virial theorem \cite{luch90}, it can be shown that 
$\langle n 0| \sqrt{\vec p\,^2} | n 0\rangle = \langle n 0| a \, r^2 | n 0\rangle$ where $| n 0\rangle$ is a
S-eigenstate. Moreover, all powers of $p$ can be computed exactly \cite{sema10}. So, we have:
\begin{align}
\label{msqrtp2}
\langle n 0|\sqrt{\vec p\,^2} | n 0\rangle &= \frac{E_{n0}}{3}, \\
\label{mAip4}
\langle n 0| \vec p\,^4 | n 0\rangle &= \left(\frac{a}{2}\right)^{4/3}\frac{16}{315}\left( 8 |\alpha_n|^4 + 25 |\alpha_n|\right).
\end{align}

\begin{figure}[htb]
\includegraphics*[height=5cm]{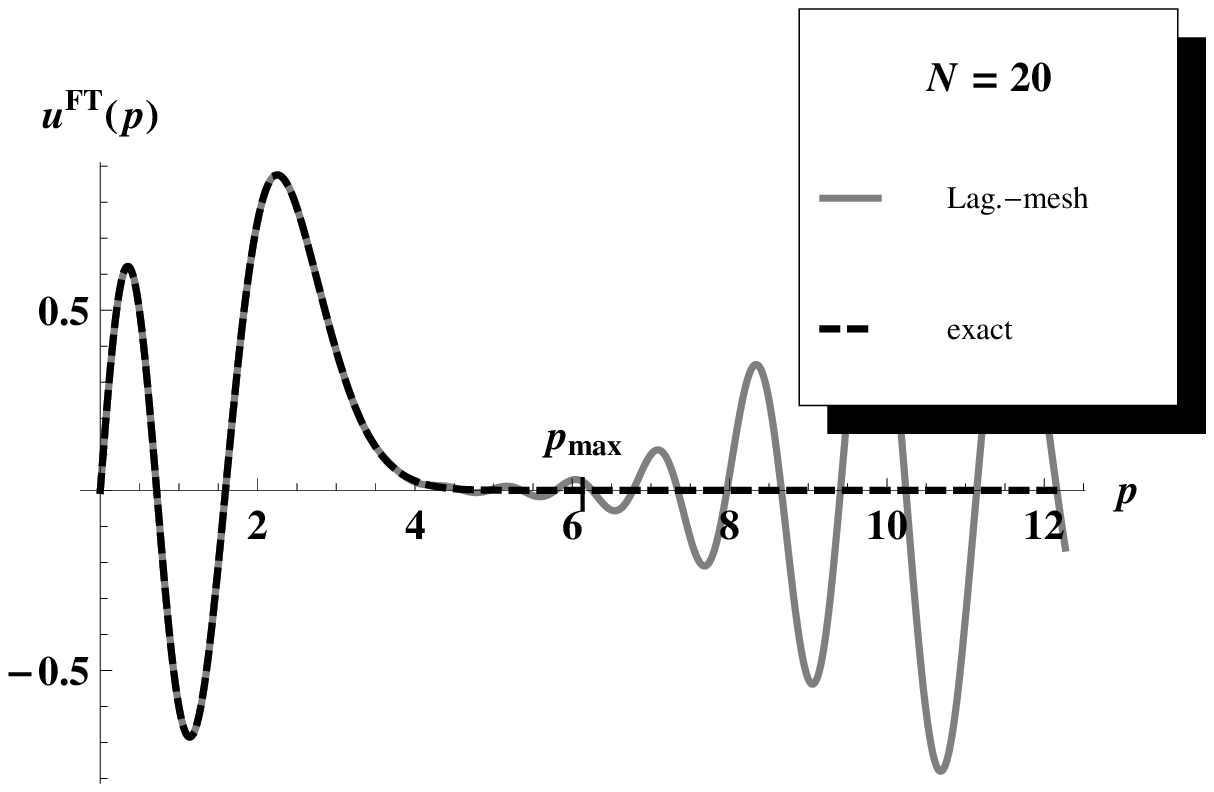}
\includegraphics*[height=5cm]{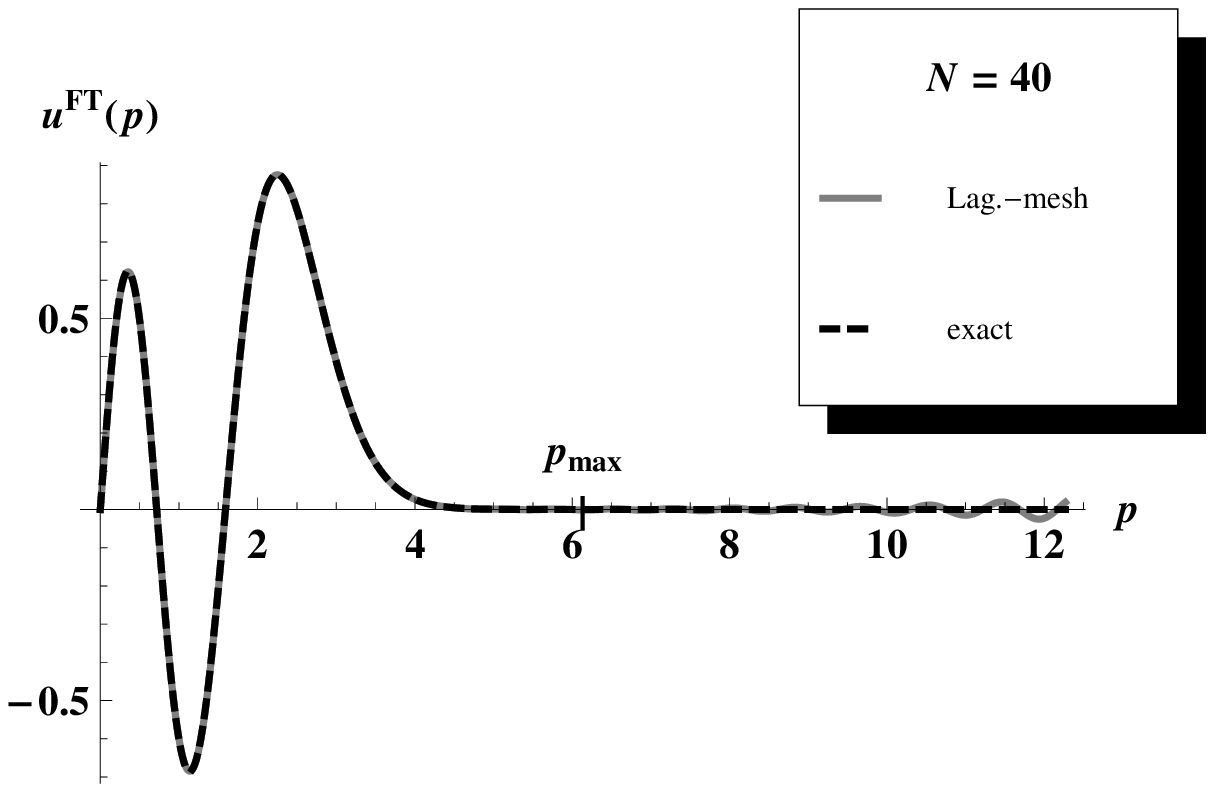}
\caption{The exact solution (\ref{foUR}) with $a=0.25$ for $n=2$ is compared with the corresponding 
approximation given by formula (\ref{FT7}) for $0 \le p \le 2 p_\textrm{max}$. 
The value of $p_\textrm{max}$ is determined with the procedure presented in Sec.~\ref{sec:scale}
with $\epsilon= 10^{-8}$.
\label{fig1}}
\end{figure}

To perform the following calculations, we have set $a=0.25$. The units of the results are given in powers 
of the unit chosen for the only energy scale of the system $a^{1/3}$.
Using the Lagrange-mesh method with $N=10$ and $\epsilon=10^{-4}$, the eigenvalues (\ref{EUR}) can already be obtained with a relative error smaller than 1\%. But, to obtain a good Fourier transform of the wavefunction, it is necessary to use more points. As we can see on Fig.~\ref{fig1}, the agreement can be very good 
for the main part of $u^\textrm{FT}(p)$. 
With $N=20$, unphysical oscillations appear just before $p_\textrm{max}$. 
With $N=40$, they develop halfway between $p_\textrm{max}$ and $2 p_\textrm{max}$.
With $N=80$ (not presented here), the asymptotic behavior is correct till $2 p_\textrm{max}$. 
In these 3 cases, for which $\epsilon=10^{-8}$, we have respectively $Q(p_\textrm{max})=0.034$, $0.0042$, 
$0.0052$. The quality factor first 
decreases rapidly due to the improvement of the wavefunction for large values of $p$, and then stabilizes
because the quality of the wavefunction stays constant in the low-$p$ part. It is possible to improve the quality
factor by decreasing the value of $\epsilon$ (increasing the value of $p_\textrm{max}$). For $N=40$, 
the value of $Q(p_\textrm{max})$ decreases from $0.015$ to $0.0020$ when $\epsilon$ varies from 
$10^{-4}$ to $10^{-12}$.

Some observables for a particular eigenstate, $l=0$ and $n=2$, computed with formula (\ref{meanK2}) are presented in Table~\ref{tab1} and compared with the exact values. Similar results are obtained for other eigenstates. A very good accuracy can be obtained with a quite small number of points. Actually, it appears that the precision does not automatically increases with $N$. On the contrary, for a given value of $\epsilon$, the accuracy is optimal for a given number of points. This behavior is typical of semirelativistic Hamiltonians. This is due to the computation of the kinetic part which requires a supplementary approximation than the use of the Gauss quadrature rule (see Sec.~\ref{ssec:kp}). Our experience is that an optimal value for an observable can be found by looking at extrema or plateau in the behavior of this observable as a function of $N$ for a given value of $\epsilon$. In the next section, we will see on an example that accuracy increases with $N$ for a nonrelativistic system. 

\begin{center}
\begin{table}[htb]
\caption{Some observables with $a=0.25$ for the eigenstate $l=0$ and $n=2$,
computed with formula (\ref{meanK2}) and compared with the exact values.
Results are given in powers of the unit for $a^{1/3}$.
\label{tab1}}
\begin{tabular}{lrlll}
\hline\hline
 & & $\left\langle \sqrt{\vec p\,^2} \right\rangle$ & $\left\langle \vec p\,^4 \right\rangle$ & $\left\langle \exp(-\vec p\,^2/a^{2/3})\right\rangle$ \\
\hline
\multicolumn{2}{c}{Exact} & 1.84019$^{(a)}$ & 24.0273$^{(b)}$ & 0.109740$^{(c)}$ \\
& & & & \\
$\epsilon=10^{-6}$ & $N=10$ & 1.84198 & 23.6260 & 0.108562 \\
                   &   $20$ & 1.84265 & 24.0735 & 0.109299 \\
                   &   $40$ & 1.84399 & 24.0982 & 0.108892 \\
& & & & \\
$\epsilon=10^{-8}$ & $N=10$ & 1.81901 & 23.6006 & 0.112181 \\
                   &   $20$ & 1.84163 & 24.0545 & 0.109512 \\
                   &   $40$ & 1.84236 & 24.0680 & 0.109359 \\
\hline\hline
\end{tabular}\\
$^{(a)}$ Computed with (\ref{msqrtp2});
$^{(b)}$ Computed with (\ref{mAip4});
$^{(c)}$ Computed with quadrature using (\ref{foUR}). 
\end{table}
\end{center}

\subsection{Hydrogen atom Hamiltonian}
\label{sec:num2}

We consider now a completely different case, the hydrogen atom: the kinematics is nonrelativistic and the Coulomb potential, $-\alpha/r$, is non-confining. The eigensolutions in the position space are well known and their Fourier transform can be expressed in term of the Appell Hypergeometric function $F_2$ \cite{niuk84}.
As these special functions are difficult and lengthy to obtain accurately, it is more convenient to work
with numerically computed eigensolutions in momentum space. Particular momentum dependent observables can be exactly computed \cite{sema10}:
\begin{align}
\label{p2coul}
\langle \vec p\,^2 \rangle &= \frac{\eta^2}{(n+l+1)^2}, \\
\label{p4coul}
\langle \vec p\,^4 \rangle &= \eta^4\frac{8 n+2 l+5}{(2 l+1)(n+l+1)^4},
\end{align}
where $\eta = \mu\,\alpha$, with $\mu$ the reduced mass.

To perform the following calculations, we have set $m_1=940$~MeV, $m_2=511$~KeV, $\alpha=1/137$. The units of the results are given in powers of keV. Some observables for a particular eigenstate, $l=1$ and $n=1$, computed with formula (\ref{meanK2}) are presented in Table~\ref{tab2} and compared with the exact values. Similar results are obtained for other eigenstates. Again, a very good accuracy can be obtained with a quite small number of points. This time, accuracy always increases with $N$ for a given value of $\epsilon$, as already found in previous studies \cite{VMB93,baye02}.

\begin{center}
\begin{table}[ht]
\caption{Some observables for the hydrogen atom eigenstate $l=1$ and $n=1$,
computed with formula (\ref{meanK2}) and compared with the exact values. 
Results are given in powers of keV.
\label{tab2}}
\begin{tabular}{lrlll}
\hline\hline
 & & $\left\langle \vec p\,^2 \right\rangle$ & $\left\langle \vec p\,^4 \right\rangle$ & $\left\langle \exp(-p/\eta)\right\rangle$ \\
\hline
\multicolumn{2}{c}{Exact} & 1.54414$^{(a)}$ & 11.9218$^{(b)}$ & 0.786997$^{(c)}$ \\
& & & & \\
$\epsilon=10^{-6}$ & $N=10$ & 1.54417 & 11.9225 & 0.787043 \\
                   &   $20$ & 1.54414 & 11.9218 & 0.786995 \\
                   &   $40$ & 1.54414 & 11.9218 & 0.786994 \\
& & & & \\
$\epsilon=10^{-8}$ & $N=10$ & 1.54711 & 11.9471 & 0.787255 \\
                   &   $20$ & 1.54414 & 11.9218 & 0.786997 \\
                   &   $40$ & 1.54414 & 11.9218 & 0.786997 \\
\hline\hline
\end{tabular}\\
$^{(a)}$ Computed with (\ref{p2coul});$^{(b)}$ Computed with (\ref{p4coul});
$^{(c)}$ Computed with quadrature of the numerical Fourier transform of the wavefunction
in position space. 
\end{table}
\end{center}

\begin{figure}[htb]
\includegraphics*[height=5cm]{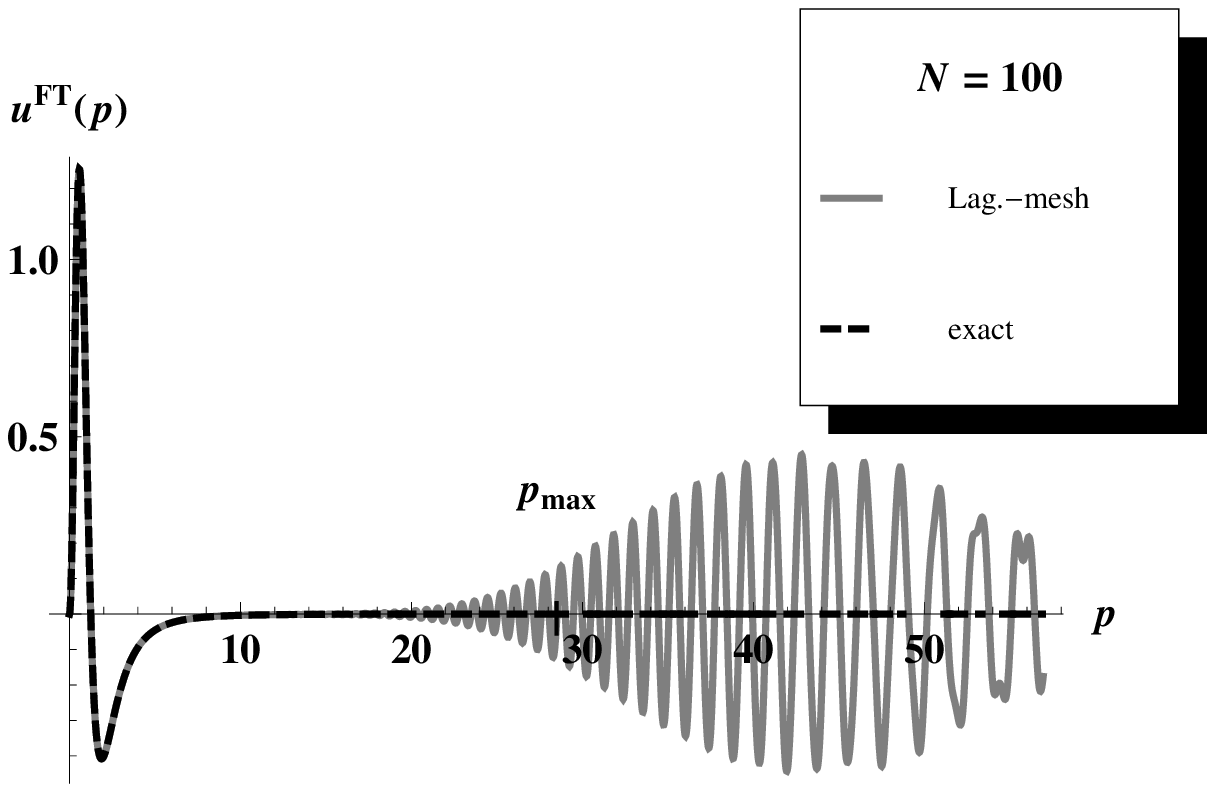}
\includegraphics*[height=5cm]{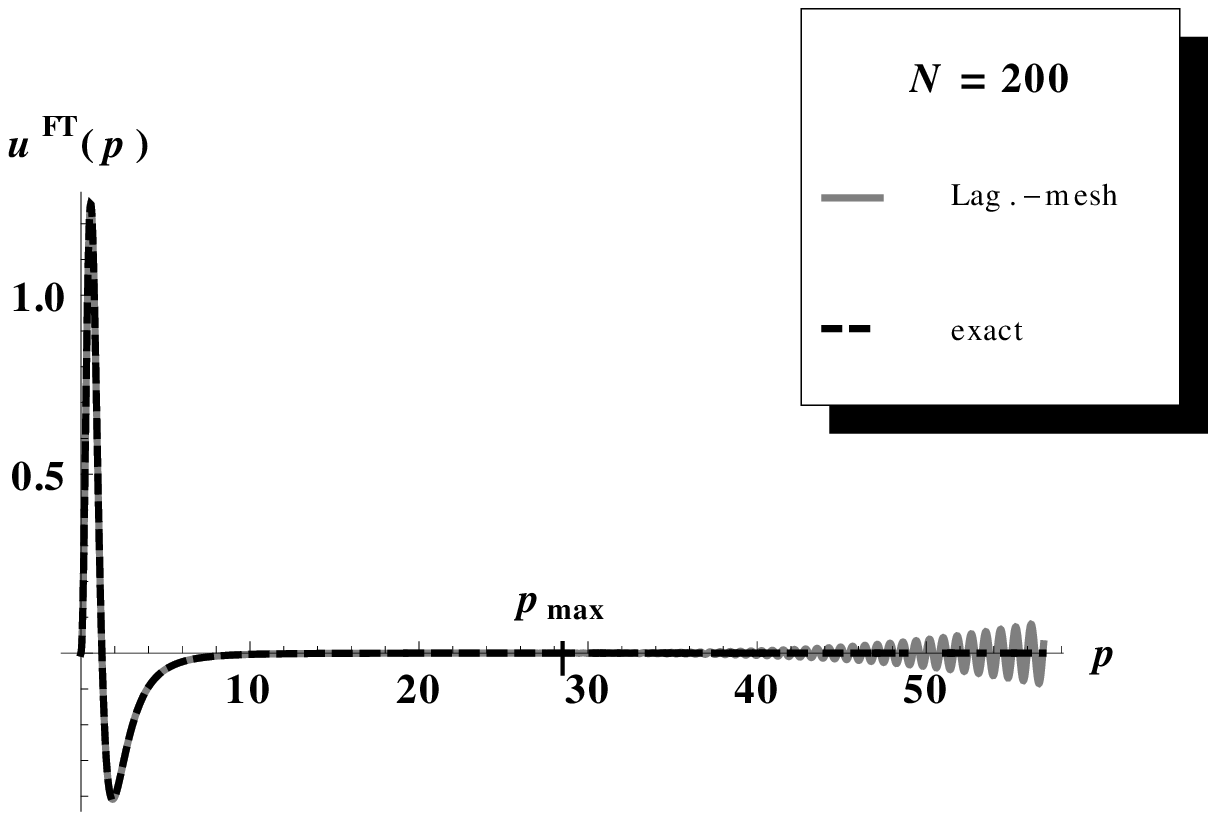}
\caption{The accurate numerically computed (exact) Fourier transform of the hydrogen atom wavefunction 
for $l=1$ and $n=1$ is compared with the corresponding 
approximation given by formula (\ref{FT7}) for $0 \le p \le 2 p_\textrm{max}$. 
The value of $p_\textrm{max}$ is determined with the procedure presented in Sec.~\ref{sec:scale}
with $\epsilon= 10^{-6}$.
\label{fig2}}
\end{figure}

A good Fourier transform of the main part of the wavefunction $u^\textrm{FT}(p)$ can be obtained with a small number of points, around $N=20$-40. But, to obtain a good asymptotic tail, it is necessary to use more points, as we can see on Fig.~\ref{fig2}.  
With $N=100$, unphysical oscillations appear before $p_\textrm{max}$. 
With $N=200$, they develop halfway between $p_\textrm{max}$ and $2 p_\textrm{max}$.
For $\epsilon=10^{-6}$, we have respectively $Q(p_\textrm{max})=0.504$, $0.097$, 
$0.00028$, for $N=50$, 100, 200. Nevertheless, the quality factor $Q(p_*)$ can be as small as $10^{-6}$ if $p_*$ is in the main part of the wavefunction. It is also possible to improve the quality
factor by decreasing the value of $\epsilon$ (increasing the value of $p_\textrm{max}$).

\section{Concluding remarks}
\label{sec:crem}

The Lagrange-mesh method is a procedure to compute eigenvalues and eigenfunctions of 
quantum equations. It is very simple to implement and can yield very accurate results for a lot
of observables, specially for nonrelativistic kinematics. At the origin, the method has been 
developed in the position space since the evaluation of potential matrix elements requires only 
the computation of the interaction at some mesh points. This is due to the use of a Gauss quadrature
rule with the fact that the basis functions satisfy the Lagrange conditions, that is to say they
vanish at all mesh points except one. Using this very special property, we have shown that 
the computation of the wavefunction in the momentum space by the Fourier transform of the wavefunction in the position space can be easily performed with a very good accuracy. Moreover, mean values of momentum dependent operators can also be easily and accurately calculated using a technique similar to the one used to compute the 
semirelativistic kinetic matrix elements. This shows again the great efficiency of
the Lagrange-mesh method which can yield very accurate results for a minimal computational 
effort. We can wonder if this technique could also be used directly in the momentum space,
for instance in the case where the interaction is only known as a function of the relative momentum.
This question will be addressed in a subsequent paper.

\section*{Acknowledgments}
C. S. would thank the F.R.S.-FNRS for the financial support. The authors are grateful to 
Fabien Buisseret for helpful suggestions.


\begin{thebibliography}{99}
\bibitem{BH-86} D.~Baye and P.-H.~Heenen, 
J. Phys. A {\bf 19}, 2041 (1986).
\bibitem{VMB93} M.~Vincke, L.~Malegat, and D.~Baye, 
J. Phys. B {\bf 26}, 811 (1993).
\bibitem{Ba-95} D.~Baye, J. Phys. B {\bf 28}, 4399 (1995).
\bibitem{Baye06} D.~Baye, Phys. Stat. Sol. (b) {\bf 243}, 1095 (2006). 
\bibitem{baye08} D. Baye and K. D. Sen, Phys. Rev. E {\bf 78}, 026701 (2008). 
\bibitem{sema01} C. Semay, D. Baye, M. Hesse, and B. Silvestre-Brac, 
Phys. Rev. E {\bf 64}, 016703 (2001).
\bibitem{brau02} F. Brau and C. Semay, 
J. Phys. G: Nucl. Part. Phys. {\bf 28}, 2771 (2002).
\bibitem{buis05} F. Buisseret and C. Semay, 
Phys. Rev. E {\bf 71}, 026705 (2005). 
\bibitem{lag1} F. Buisseret and C. Semay, 
Phys. Rev. E \textbf{75}, 026705 (2007). 
\bibitem{baye02} D. Baye, M. Hesse, and M. Vincke, 
Phys. Rev. E {\bf 65}, 026701 (2002). 
\bibitem{baye99} D. Baye and M. Vincke, Phys. Rev. E {\bf 59}, 7195 (1999).
\bibitem{hess99} M. Hesse and D. Baye, J. Phys. B {\bf 32}, 5605 (1999).
\bibitem{numrec} W. H. Press, S. A. Teukolsky, W. T. Vetterling, and B. P. Flannery, 
\textit{Numerical Recipes} (Cambridge University Press, 2007).
\bibitem{golu69} G. H. Golub and J. H. Welsch, Math. Comput. {\bf 23}, 221 (1969).
\bibitem{fulc94} L. P. Fulcher, Phys. Rev. D {\bf 50}, 447 (1994).
\bibitem{var} D. A. Varshalovich, A. N. Moskalev, and V. K. Khersonskii, 
\textit{Quantum Theory of Angular Momentum} (World Scientific, Singapore, 1988).
\bibitem{abra65} M. Abramowitz and I. A. Stegun, 
\textit{Handbook of Mathematical Functions} (Dover, New York, 1965).
\bibitem{brau98} F. Brau and C. Semay, 
J. Comput. Phys. {\bf 139}, 127 (1998).
\bibitem{LambW} R. M. Corless, G. H. Gonnet, D. E. G. Hare, D. J. Jeffrey, and D. E. Knuth,
Adv. Comput. Math., 329 (1996).  
\bibitem{luch91} W. Lucha, F. F. Sch\"oberl, and D. Gromes, Phys. Rep. {\bf 200}, 127 (1991).
\bibitem{Sem03} C. Semay, B. Silvestre-Brac, and I. M. Narodetskii,
Phys. Rev. D {\bf 69}, 014003 (2004).
\bibitem{luch90} W. Lucha, Mod. Phys. Lett. A {\bf 5}, 2473 (1990). 
\bibitem{sema10} C. Semay and B. Silvestre-Brac, J. Phys. A: Math. Theor. {\bf 43}, 265302 (2010).
\bibitem{niuk84} A.W. Niukkanen, Int. J. Quantum. Chem. {\bf 25}, 941 (1984).
\end{thebibliography}
\end{document}